\documentstyle[12pt,epsf]{article}
\def\lsim{\mathrel{\lower2.5pt\vbox{\lineskip=0pt\baselineskip=0pt
           \hbox{$<$}\hbox{$\sim$}}}}
\def\gsim{\mathrel{\lower2.5pt\vbox{\lineskip=0pt\baselineskip=0pt
           \hbox{$>$}\hbox{$\sim$}}}}
\begin{document}
\setlength{\baselineskip}{8mm}
\begin{titlepage}
\begin{flushright}
\begin{tabular}{c c}
& {\normalsize  August 1996}
\end{tabular}
\end{flushright}
\vspace{5mm}
\begin{center}
{\large \bf Explicit $CP$ Violation in the Higgs Sector of \\
the Next-to-Minimal Supersymmetric Standard Model } \\
\vspace{15mm} 
Naoyuki Haba\footnote{E-mail:\ haba@eken.phys.nagoya-u.ac.jp} \\ 

{\it 
Department of Physics, Nagoya University \\
           Nagoya, JAPAN 464-01 \\
}
\end{center}

\vspace{10mm}


\begin{abstract}

We analyze 
explicit $CP$ violation in the Higgs sector of 
the next-to-minimal supersymmetric standard model 
which contains an additional gauge singlet field $N$. 
It is shown that there is no mixing among 
scalar and pseudoscalar Higgs fields in the two 
Higgs doublets, and 
scalar-pseudoscalar mixings could exist 
between two Higgs doublets and the singlet $N$, 
and between $N$ itself. 
$CP$ symmetry is conserved in the extreme limits of 
$\langle N \rangle \gg v$, $\langle N \rangle \ll v$, 
and $\tan \beta \gg 1$. 
In the region of $\langle N \rangle = O(v)$ 
and $\tan \beta = O(1)$, 
large scalar-pseudoscalar mixings are realized, 
and this effect can reduce 
the lightest Higgs mass. 
The mass difference between no mixing and 
mixing case is about 
$10 \sim 30$ GeV. 
The neutron electric dipole moment 
in this model is 
consistent with the present experimental upper limit 
provided that 
squark and gaugino masses are heavy enough of $O(1)$ TeV.

\end{abstract}
\end{titlepage}

%
\section{Introduction}

The origin of $CP$ violation is one of the most exciting 
topics in the present particle physics. 
In the standard model (SM), 
$CP$ phase exists in 
the Kobayashi-Maskawa (KM) matrix\cite{KM}, and 
$CP$ is automatically conserved in the Higgs sector. 
However, in multi-Higgs models, 
$CP$ could be violated explicitly or spontaneously 
in the Higgs sector\cite{THDM}\cite{THDM2}. 
The natural model containing two Higgs doublets is 
the minimal supersymmetric standard model (MSSM). 
The Higgs potential of the MSSM could break 
$CP$ symmetry explicitly or spontaneously 
taking account of radiative corrections. 
However the radiative violation leads to the fact 
that explicit $CP$ violation through 
the Higgs sector in the MSSM is too small 
to have any significant 
phenomenological implications 
and spontaneous $CP$ violation 
in the MSSM also requires the existence 
of a light Higgs "pseudoscalar" 
which is inconsistent with 
experiments\cite{MAEKAWA}. 
So we should extend the model in order to have 
significant $CP$ violation effects 
in the Higgs sector in SUSY models. 
The minimal extension is adding the gauge singlet 
field $N$ to the MSSM, 
which is so-called next-to-minimal 
supersymmetric standard model (NMSSM)\cite{NMSSM}. 
\par
In this paper we concentrate on 
explicit $CP$ violation scenario in the NMSSM. 
Spontaneous $CP$ violation in the NMSSM is 
discussed in Refs.\cite{BABU}\cite{HABA}. 
Explicit $CP$ violation in the Higgs sector 
is possible by complex couplings 
at the tree level Lagrangian. 
As shown later, 
scalars and a pseudoscalar of two Higgs doublets 
do not mix and 
scalar-pseudoscalar mixings only exist 
among Higgs doublets and the singlet 
and among the singlet itself at the tree level. 
These results can be derived by using 
the stationary conditions of phases 
and are completely different from the results obtained in 
Ref.\cite{MATSUDA}. 
These mixings vanish in the limits 
of $x \gg v$, $x \ll v$, and $\tan \beta \gg 1$, 
where $v$ is the electroweak scale and 
$x$ is the vacuum expectation value (VEV) 
of the singlet field $N$. 
In the case of $x = O(v)$ and $\tan \beta = O(1)$, 
large mixings among scalars and pseudoscalars are 
realized in the Higgs sector. 
Available parameters are restricted by experiments and 
the lightest Higgs mass becomes smaller than the mass 
of the no mixing case 
by the effect of scalar-pseudoscalar mixings. 
The predicted neutron electric dipole moment (NEDM) 
could be consistent with the present experiment 
provided that squark and gaugino masses are heavy enough 
of $O(1)$ TeV. 
\par
Section 2 is devoted to explicit 
$CP$ violation of the Higgs sector in the NMSSM. 
In section 3, we discuss the numerical result by using 
recent experimental constraints. 
Section 4 gives summary and discussion. 
%

%
\section{Explicit $CP$ Violation in the NMSSM Higgs Sector}

The superpotential of the NMSSM which contains 
only top Yukawa coupling for the 
Yukawa sector is given by 
\begin{equation}
\label{WNMSSM}
W = h_tQH_2T^c + \lambda N H_1 H_2 - {k \over 3} N^3 . 
\end{equation}
$H_1$ and $H_2$ are Higgs doublet fields denoted as 
\begin{equation}
H_1=\left(
\begin{array}{c}
H_1^{0} \\
H_1^- \\
\end{array}
\right) , \qquad
H_2=\left(
\begin{array}{c}
H_2^+ \\
H_2^0 \\
\end{array}
\right),
\end{equation}
with 
\begin{equation}
  H_1H_2=H_1^0H_2^0-H_1^-H_2^+. 
\end{equation}
The third generation quark doublet superfield is 
denoted as $Q$, and 
$T^c$ is the right-handed top quark superfield. 
The coupling $h_t$ is the top Yukawa coupling constant. 
One pseudoscalar which is the mixing state of $H_1$ and 
$H_2$ is the Goldstone boson absorbed by $Z$ boson. 
There are three neutral scalars, two 
neutral pseudoscalars, and one charged Higgs particle 
as the physical particles in the limit 
of no $CP$ violation in the Higgs sector of the NMSSM. 
\par
The Higgs potential of the NMSSM derived from 
Eq.(\ref{WNMSSM}) is given by 
\begin{equation}
\label{nmssmpot}
V = V_{\rm no phase} + V_{\rm phase} +  V_{\rm top} \ , 
\end{equation}
\begin{eqnarray}
\label{phase1}
 V_{\rm no phase} &=& m_{H_1}^2 |H_1|^2 + 
           m_{H_2}^2 |H_2|^2 + m_N^2 |N|^2 
                                        \nonumber \\ 
                  & & + {g_1^2 + g_2^2 \over 8}
                (|H_1|^2-|H_2|^2)^2 + {g_2^2 \over 2} 
  (|H_1|^2 |H_2|^2 - |H_1 H_2|^2)         \\
                  & & + |\lambda|^2 [|H_1 H_2|^2 + 
                    |N|^2 ( |H_1|^2 + |H_2|^2 )] + 
                |k|^2 |N|^4  ,           \nonumber  \\
\label{phase}  
 V_{\rm phase}    &=& -(\lambda k^* H_1 H_2 N^{*2} +
           \lambda A_{\lambda} H_1 H_2 N  +   
           {k A_k \over 3} N^3 + {\rm h.c.})  \ ,  \\
\label{topeffect}
 V_{\rm top} &=& {3 \over 16 \pi^2} 
    \left[ (h_t^2 |H_2|^2 + m_{\tilde t}^2)^2 {\rm ln} 
          {(h_t^2 |H_2|^2 + m_{\tilde t}^2) \over Q^2} - 
            h_t^4 |H_2|^4 {\rm ln} 
          {h_t^2 |H_2|^2 \over Q^2} \right] \ . 
\end{eqnarray}
$V_{\rm top}$ represents top and stop one loop radiative 
corrections\cite{COLMAN} where we assume that 
the soft breaking masses satisfy 
$m_{\tilde t_L} = m_{\tilde t_R} \equiv m_{\tilde t} \gg m_t$. 
The parameters $\lambda, k, A_\lambda$, and $A_k$ 
are all complex in general. 
$CP$ phase only appears in 
$V_{\rm phase}$ and 
we can remove two complex phases by 
the field redefinition of $H_1 H_2$ and $N$\cite{MATSUDA}. 
Therefore, without loss of generality, we can take 
\begin{equation}
\lambda A_{\lambda} > 0, \;\;\;\;\; k A_k > 0. 
\end{equation}
Only one phase remains in $\lambda k^*$ denoted as 
\begin{equation}
 \lambda k^* \equiv \lambda k e^{i \phi}, 
\end{equation}
where $\lambda$ and $k$ on the right hand side are 
real and positive parameters. 
VEVs of $H_1, H_2$, and $N$ are complex in general, 
which should be determined 
by stationary conditions. 
We denote VEVs as 
\begin{equation}
\label{NMSSMtheta}
 \langle H_1 \rangle = v_1 e^{i \varphi_1}, \;\;\;\;
 \langle H_2 \rangle = v_2 e^{i \varphi_2}, \;\;\;\; 
 \langle N \rangle = x e^{i \varphi_3} ,
\end{equation}
where $v_1$ and $v_2$ are 
real and positive which satisfy 
$v \equiv \sqrt{v_1^2+v_2^2}=174$ GeV. 
Higgs fields are defined as 
\begin{eqnarray}
\label{tenkai1}
 H_1^0 &=& v_1 e^{i \varphi_1} + 
         {1 \over \sqrt{2}} e^{i \varphi_1} 
          (S_1+i\sin\beta A) ,      \nonumber \\
 H_2^0 &=& v_2 e^{i \varphi_2} + 
         {1 \over \sqrt{2}} e^{i \varphi_2} 
                   (S_2+i\cos\beta A) ,       \\
 N     &=& x e^{i \varphi_3} + 
         {1 \over \sqrt{2}} e^{i \varphi_3} (X+iY) , 
                                       \nonumber
\end{eqnarray}
where $X$ and $Y$ are the scalar and 
the pseudoscalar field of the singlet $N$, respectively, 
and $\tan \beta \equiv v_2 / v_1$. 
\par
Now we calculate mass spectra. 
There are two physical phases of VEVs, for which we take 
\begin{equation}
 \theta \equiv \varphi_1 + \varphi_2 + \varphi_3, \;\;\;\;\;
 \delta \equiv 3 \varphi_3.
\end{equation}
Stationary conditions of phases 
\begin{equation}
\label{stationaryconditionp2}
 \left.{\partial V \over \partial \theta}\right| = 0 , \;\;\;\;
 \left.{\partial V \over \partial \delta}\right| = 0 ,
\end{equation}
induce the equations 
\begin{eqnarray}
\label{Exp1}
 & & \lambda k x^2 \sin (\phi + \theta - \delta) + 
     \lambda A_{\lambda} x \sin \theta = 0  ,  \\
\label{Exp2}
 & & -3 \lambda k v_1 v_2 \sin (\phi + \theta - \delta) + 
      k A_k x \sin \delta = 0 ,     
\end{eqnarray}
respectively. 
It is noteworthy that the phase $\phi$, which induces 
explicit $CP$ violation in the Higgs sector, 
should be $0$ or $\pi$ 
{}from Eqs.(\ref{Exp1}) and (\ref{Exp2}) 
if VEVs have no phases 
($\theta = \delta = 0$). 
Therefore we can not take $\theta = \delta = 0$ 
in the case of $\phi \neq 0, \pi$ 
contrary to the results in Ref.\cite{MATSUDA}. 
Three parameters $m_{H_1}^2, m_{H_2}^2$, and $m_N^2$ 
are eliminated by three stationary conditions 
\begin{equation}
\label{mini1}
 \left.{\partial V \over \partial v_i}\right| = 0 
 \; (i=1,2), \;\;\;\;\;\; 
 \left.{\partial V \over \partial x}\right| = 0 . 
\end{equation}
Then we get $5 \times 5$ 
neutral Higgs mass matrix 
\begin{equation}
\label{mat5*5}
 M_{H^0}^2 = 
 \left(
  \begin{array}{cc}
   M^{S_1,S_2,X}_{S_1,S_2,X}    & M^{A,Y}_{S_1,S_2,X}       \\ 
                                &                           \\
   (M^{A,X}_{S_1,S_2,X})^T      & M^{A,Y}_{A,Y}       
  \end{array}
 \right) \;,
\end{equation}
where $M^{S_1,S_2,X}_{S_1,S_2,X}$, $M^{A,Y}_{S_1,S_2,X}$, and 
$M^{A,Y}_{A,Y}$ are $3 \times 3$, $3 \times 2$, and $2 \times 2$ 
submatrices, respectively. 
The matrix $M^{S_1,S_2,X}_{S_1,S_2,X}$ of 
the scalar part of $S_1, S_2$, and 
$X$ is given by 
\begin{equation}
\label{mat3*3}
 M^{S_1,S_2,X}_{S_1,S_2,X} = 
 \left(
  \begin{array}{lll}
   \overline{g}^2 v^2 \cos^2 \beta  & 
        (\lambda^2 - \overline{g}^2/2) v^2 \sin 2 \beta &
                              2 \lambda^2 v x \cos \beta \\
   + \lambda x A_{\sigma_1} \tan \beta  & 
            -\lambda x A_{\sigma_1} &  
                -\lambda v \sin \beta A_{\sigma_2} \\
                         &  &       \\
   (\lambda^2 - \overline{g}^2/2) v^2 \sin 2 \beta & 
            (\overline{g}^2+ \Delta) v^2 \sin^2 \beta &
                             2 \lambda^2 v x \sin \beta  \\
   - \lambda x A_{\sigma_1}            &
            + \lambda x A_{\sigma_1}/ \tan \beta &
                     -\lambda v \cos \beta A_{\sigma_2} \\ 
                                     &  &       \\
   2 \lambda^2 v x \cos \beta         &
          2 \lambda^2 v x \cos \beta        &
   {\lambda v^2 \over 2 x} A_{\lambda}
                \cos \theta \sin 2 \beta \\
   - \lambda v \sin \beta A_{\sigma_2} &
               - \lambda v \cos \beta A_{\sigma_2} &
                   - A_k k x \cos \delta + 4 k^2 x^2  
  \end{array}
 \right) \;,
\end{equation}
where $A_{\sigma_1} \equiv A_{\lambda}\cos \theta + 
k x \cos (\phi + \theta - \delta)$ and 
$A_{\sigma_2} \equiv A_{\lambda}\cos \theta + 
2 k x \cos (\phi + \theta - \delta)$. 
The $\Delta$ is derived from Eq.(\ref{topeffect}) 
and given as 
\begin{equation}
\label{DELTA} 
\Delta \equiv {3 h_t^4 \over 4 \pi^2} \; {\rm ln}\: 
              {m_t^2 + m_{\tilde t}{}^2 \over m_t^2} \;. 
\end{equation}
The matrix $M^{A,Y}_{A,Y}$ of the pseudoscalar part of $A$ and 
$Y$ is given by 
\begin{equation}
\label{mat2*2}
 M^{A,Y}_{A,Y} = 
 \left(
  \begin{array}{cc}
   2 \lambda x A_{\sigma_1}/ \sin 2 \beta    &  
                                 \lambda v A_{\sigma}'   \\ 
                                       &       \\
   \lambda v A_{\sigma}'               & 
    {\lambda v^2 \over 2 x}
      A_{\lambda} \sin 2 \beta \cos \theta + 
            3 A_k k x \cos \delta \\
           & + 2 \lambda k v^2 \sin 2 \beta 
                  \cos (\phi + \theta - \delta)
  \end{array}
 \right) \;,
\end{equation}
where we define $A_{\sigma}' \equiv 
A_{\lambda}\cos \theta - 
2 k x \cos (\phi + \theta - \delta)$. 
The scalar-pseudoscalar mixing matrix 
$M^{A,Y}_{S_1,S_2,X}$ is given by 
\begin{equation}
\label{mat2*3}
 M^{A,Y}_{S_1,S_2,X} = 
 \left(
  \begin{array}{cc}
      0    & -3 \lambda k v x \sin \beta 
               \sin (\phi + \theta - \delta)   \\
      0    & -3 \lambda k v x \cos \beta 
               \sin (\phi + \theta - \delta) \\
   \lambda k v x \sin (\phi + \theta - \delta)  
           & -2 \lambda k v^2 \sin 2 \beta 
               \sin (\phi + \theta - \delta) 
  \end{array}
 \right) \;.
\end{equation}
The mixing with scalar components $S_1, S_2$ 
and the pseudoscalar component $A$ always vanish 
even if $CP$ phase $\phi$ takes non-zero values. 
It is noted that scalar-pseudoscalar mixings 
among two Higgs doublets at the tree level 
does not exist contrary to the Ref.\cite{MATSUDA}. 
In the case of $\sin (\phi + \theta - \delta)= 0$, 
all mixing with scalars and pseudoscalars 
vanish. 
\par 
The physical charged Higgs field is defined as 
$C^+ \equiv \cos \beta H^+ + \sin \beta H^{-*}$ and 
its mass is given by 
\begin{equation}
 m_C^2 = m_W^2 - 
     \lambda v^2 + {2 \lambda A_{\sigma_1} 
     x \over \sin 2 \beta}. 
\end{equation}
\par
Now we show parameters $x$ and $\tan \beta$ dependence of 
the magnitude of scalar and pseudoscalar 
mixing. 
As for $x$, we consider three special 
limiting cases\cite{MATSUDA}\cite{ELLIS}; 
{\bf (1)} $x \gg v_1, v_2$ with $\lambda$ and $k$ fixed, 
{\bf (2)} $x \gg v_1, v_2$ with $\lambda x$ and $k x$ fixed, and 
{\bf (3)} $x \ll v_1, v_2$. 
And also we consider the case {\bf (4)} with 
the limit of $\tan \beta \gg 1$. 
\begin{description}
\item[(1)]
Limits of $x \gg v_1, v_2$ ($\lambda$ and $k$ fixed); \\
In this limit with $A_{\lambda}, A_k = O(x)$, 
$\sin \delta$ goes to zero from Eq.(\ref{Exp2}). 
$M_X^X$ and $M_Y^Y$ elements are of $O(x^2)$ 
because of $\cos \delta \simeq 1$ 
and mixing components with $S_1, S_2, A$ and $X, Y$ are of 
$O(v x)$. 
Then the mixing angles of 
scalars and pseudoscalars become 
$O(v/x)$. 
Therefore $X$ and $Y$ are heavy enough to 
decouple from $S_1, S_2$, and $A$, and 
it is enough to consider the 
$3 \times 3$ submatrix of $S_1 - S_2 - A$ to 
estimate $CP$ violation effects. 
We know that $S_1, S_2$, and $A$ do not mix from 
Eq.(\ref{mat2*3}). 
Then $CP$ violation in the Higgs sector vanishes 
in this limit. 
\item[(2)]
Limits of $x \gg v_1, v_2$ ($\lambda x$ and $k x$ fixed); \\
This limit with 
$A_{\lambda}, A_k = O(\lambda v)$ 
reduces the NMSSM to the MSSM. 
We can easily show that 
all scalar-pseudoscalar mixing elements 
in Eq.(\ref{mat2*3}) vanish. 
Then 
$CP$ symmetry in the Higgs sector restores 
in this limit. 
\item[(3)]
Limits of $x \ll v_1, v_2$; \\
In this limit with $A_{\lambda}, A_k = O(v)$, 
$\sin (\phi + \theta - \delta)$ goes to zero from 
Eq.(\ref{Exp2}). 
Then all components of $M^{A,Y}_{S_1,S_2,X}$ vanish and 
the Higgs potential has $CP$ symmetry in this limit. 
\item[(4)]
Limits of $\tan \beta \gg 1$; \\
In the large $\tan \beta$ limit, 
$S_2$-$Y$ and $X$-$Y$ components in the 
$M_{S_1,S_2,X}^{A,Y}$ vanish. 
Then it is enough to consider the $S_1$-$Y$ and 
$S_2$-$X$-$A$ submatrices. 
These are 
\begin{equation}
\label{matS1Y}
 M^{S_1,Y}_{S_1,Y} = 
 \left(
  \begin{array}{cc}
   \lambda x A_{\sigma_1} \tan \beta & 
      -3 \lambda k v x \sin (\phi + \theta - \delta) \\
   -3 \lambda k v x \sin (\phi + \theta - \delta)  
              & 3 A_k k x \cos \delta   
  \end{array}
 \right) \; ,
\end{equation}
and 
\begin{equation}
\label{matS2XA}
 M^{S_2,X,A}_{S_2,X,A} = 
 \left(
  \begin{array}{ccc}
   (\overline{g^2} + \Delta) v^2  &  
                      2 \lambda^2 v x  &  0  \\
   2 \lambda^2 v x                &  
     -A_k k x \cos \delta + 4 k^2 x^2 &  
         2 \lambda k v x \sin (\phi + \theta - \delta) \\
                              0   &  
         2 \lambda k v x \sin (\phi + \theta - \delta) & 
            2 \lambda x A_{\sigma_1}/ \sin 2 \beta \\
  \end{array}
 \right) \;,
\end{equation}
respectively. 
Only $S_1$-$S_1$ and $A$-$A$ components 
are dominant in each matrix 
unless $A_{\sigma_1}=0$. 
Then scalar-pseudoscalar mixings 
become negligibly small and 
$CP$ violation in the Higgs sector vanishes 
in this limit. 
If $A_{\sigma_1}=0$, 
the phase $\phi$ should be $0$ or $\pi$ from 
Eq.(\ref{Exp1}) since $\delta$ goes to 
be $0$ or $\pi$ from Eq.(\ref{Exp2}). 
It is not the case of explicit $CP$ violation 
of the Higgs sector in the NMSSM. 
\end{description}
\par
In the above various limits of {\bf (1)}$\sim${\bf (4)}, 
$CP$ symmetry is approximately conserved in the Higgs sector. 
As for the region of $x = O(v)$ and 
$\tan \beta = O(1)$, 
large scalar-pseudoscalar mixings are realized. 
We will discuss this case 
numerically in the next section.

%
\section{Numerical Analysis of Explicit $CP$ Violation}

In this section we show numerical 
examples to realize the large $CP$ violation. 
We discuss the case of $x = O(v)$ and 
$\tan \beta = O(1)$. 
\par
In Fig.1 we show the experimentally allowed region 
in the $\cos \phi - \lambda$ plane for fixed values 
of other parameters. 
The values of parameters are 
\begin{eqnarray}
\label{25}
 & &  k = 0.1,\;\;\;\;\; \tan \beta = 2, \;\;\;\;\; 
 m_{\tilde t} = 3 \;{\rm TeV}, \nonumber \\
 & &  A_k = v, \;\;\;\;\;  A_{\lambda} = v/2, 
 \;\;\;\;\; x = 5 \; v.
\end{eqnarray}
Here we consider the following experimental 
constraints\cite{HABA}; 
\begin{description}
\item[(A)]
The lightest Higgs boson $(h_1)$ and 
the second lightest Higgs boson $(h_2)$ 
have not been observed 
in the decay of $Z$. 
So the condition of 
$$
m_{h_1}+m_{h_2} > m_Z
$$
should be satisfied kinematically, or 
in the case of the sum of 
$m_{h_1}$ and $m_{h_2}$ is smaller than $m_Z$, 
the branching ratio 
$B(Z \rightarrow h_1h_2)$ should be less than 
$10^{-7}$\cite{PDG}. 
The decay rate is 
\begin{equation}
 \Gamma(Z \rightarrow h_1h_2) = 
  {M_Z \over 16\pi}g^2_{Zh_1h_2}\lambda^{3 \over 2}
  (1, x_{1},x_{2}),
\end{equation}
where $\lambda(x,y,z) \equiv x^2+y^2+z^2-2xy-2yz-2zx$ 
and $x_{i} \equiv m^2_{h_i}/M^2_Z$. 
The coupling of $g_{Zh_1h_2}$ is 
defined as 
\begin{equation}
 g_{Zh_1h_2} \equiv g_2 
       [ \cos \beta (a_{1 S_2}a_{2 A}-a_{2 S_2}a_{1 A}) - 
         \sin \beta (a_{1 S_1}a_{2 A}-a_{2 S_1}a_{1 A}) ],
\end{equation}
where $a_{iJ}$ is the ratio of the field $J$ component 
in $h_i$ $(i=1,2)$. 
\item[(B)]
The Higgs $h_i$ has not been 
observed in the decay $Z \rightarrow h_i+Z^* 
\rightarrow h_i +l^+l^-$
\cite{ALEPH}.
So $B(Z \rightarrow h_i l^+l^-)$ should be smaller than 
$1.3 \times 10^{-7}$\cite{PDG}. 
In the case $m_{h_1}$ and/or $m_{h_2}$ being 
smaller than $M_Z$, the decay rate becomes 
\begin{eqnarray}
 \Gamma(Z \rightarrow h_i l^+l^-) &=& 
  {1 \over 96\pi^3}{g^2_{ZZh_i}g^2_{Zl^+l^-} \over M_Z}
              (|C_L|^2+|C_R|^2)       \nonumber\\
       & &  
             \int_{\rho_i}^{1+\rho_i^2 \over 2}
    {1+\rho_i^2-2x \over (\rho_i^2-2x)^2+\Gamma_Z^2/(4M_Z^2)}
     (x^2-\rho_i^2)^{1/2} dx, 
\end{eqnarray}          
where 
$\rho_i=m_{h_i}/M_Z$, $x=E_{h_i}/M_Z$, 
$g_{Zl^+l^-}=2e/\sin 2\theta_W$,
$C_L=-{1 \over 2} + \sin^2 \theta_W$ and $C_R=\sin^2\theta_W$. 
The coupling constant 
$g_{ZZh_i}$ is given by 
\begin{equation}
\label{ZZh}
 g_{ZZh_i}\equiv{g_2 \over 2 \cos \theta_W}
            M_Z \cos \beta (a_{i S_1}+
                   a_{i S_2} \tan \beta) .
\end{equation}
\item[(C)]
The chargino ${\tilde \chi}^{\pm}$ has not been observed 
in the decay $Z \rightarrow {\tilde \chi}^{+} 
{\tilde \chi}^{-}$, the mass of chargino 
$M_{{\tilde \chi}^{\pm}}$must be 
larger than 45.2 GeV\cite{PDG}. 
\end{description}
\vspace{5mm}
The $\mu$ term in the MSSM is represented as 
$\lambda x$ in the NMSSM, 
so the chargino mass becomes small 
as $\lambda$ becomes small. 
With the wino soft breaking mass $M_2$ being 1 TeV, 
the dashed boundary in the figure 
is derived from the constraint {\bf (C)}. 
\par
In Fig.2 the experimentally allowed region of 
$\lambda - \tan \beta$ plane is shown at 
$\cos \phi = 0$, 
where we take other parameters as the same 
in Eq.(\ref{25}). 
The allowed region of $\lambda$ is about 
$0.06 \sim 0.24$ which has small 
dependence on $\tan \beta$. 
We only consider the region of $\tan \beta \geq 1$ 
in this paper\cite{NOJIRI}. 
\par
In Fig.3 the mass of the lightest Higgs particle 
versus $\lambda$ is shown. 
The region of $\lambda$ is limited from constraints 
{\bf (A)}$\sim${\bf (C)}. 
$CP$ violation in the Higgs sector 
requires scalar-pseudoscalar mixing, 
which reduces the magnitude of the mass 
up to $10 \sim 30$ GeV. 
The qualitative result is not changed 
in the region of $m_{\tilde t} = 
1 \sim 3$ TeV. 
\par
Now we discuss the NEDM. 
The phases which contribute to 
the NEDM at one loop level are induced 
{}from the chargino, the neutralino, 
and the squark mass matrices. 
As following the analysis by Kizukuri and Oshimo\cite{OSHIMO}, 
the chargino diagram gives the dominant contribution to 
the NEDM under the 
GUT relation of gaugino soft masses. 
The NEDM is proportional to $\lambda$ and 
$\sin (\theta + \eta)$, 
where $\eta$ is the phase of $M_2$. 
Then the maximal NEDM 
occurs at $\lambda = 0.27$ and 
$\sin (\theta + \eta) = 1$. 
The calculated value of the NEDM is 
$3.0 \times 10^{-27}$ e$\cdot$cm 
with $M_2 = 1$ TeV and 
squark masses $m_{{\tilde u},{\tilde d}} = 3$ TeV. 
This value is about two orders smaller than 
the experimental upper bound\cite{PDG}. 
Squark masses should be of $O(1)$ TeV 
to retain consistency with the EDM experiment. 
{}For example, in the case of 
$\lambda = 0.23$ which satisfy 
constraints {\bf (A)}$\sim${\bf (C)}, 
squark masses must be larger than 1.4 TeV.

%
%
\section{Summary and Discussion}

We have studied explicit $CP$ violation in the Higgs 
sector of the NMSSM. 
Mixings with scalars and a pseudoscalar of two Higgs doublets 
do not exist by the stationary conditions of phases. 
Scalar-pseudoscalar mixings only exist 
through the singlet field $N$. 
This situation is not changed drastically 
by radiative corrections because 
loop effects, which contribute scalar-pseudoscalar mixings, 
are negligibly small as same as the 
case of explicit $CP$ violation in the MSSM. 
These mixings would vanish in the limits 
of $x \gg v$, $x \ll v$, and $\tan \beta \gg 1$. 
In the region of $x = O(v)$ and $\tan \beta = O(1)$, 
$CP$ violating 
effects are realized in the Higgs sector. 
The lightest Higgs mass is reduced about 
$10 \sim 30$ GeV 
by the effect of scalar-pseudoscalar mixings. 
The value of the NEDM predicted by this model 
is consistent with the experiment 
provided that squark and gaugino masses are heavy enough 
of $O(1)$ TeV.

\vskip 1 cm
\noindent
{\bf Acknowledgements}\par
I would like to thank Professor 
M. Matsuda for 
useful discussions and careful reading of manuscripts. 
I am also grateful to Professor A. I. Sanda and 
Professor M. Tanimoto for useful comments and discussions.


%
%
\newpage

\begin{center}
{\bf Figure Captions} \\
\end{center}
{\bf Fig.1}\qquad
The allowed region in the $\cos \phi - \lambda$ plane 
for $k = 0.1, \tan \beta = 2, m_{\tilde t} = 3$ TeV, 
$A_k = v$, $A_{\lambda}= v/2$, and $x = 5 v$. 
The solid boundary corresponds to constraints {\bf (A)} 
and {\bf (B)}. 
The dashed boundary 
corresponds to the constraint {\bf (C)}. \\
{\bf Fig.2} \qquad
The allowed region in the $\tan \beta - \lambda$ plane 
at $\cos \phi = 0$. Other parameters are given in Eq.(\ref{25}) 
and the boundaries are obtained by the same 
constraints in Fig.1. \\
{\bf Fig.3} \qquad
$\lambda$ dependence of 
the predicted lightest Higgs particle mass; 
short-dashed line: $\cos \phi = 1$ (no mixing case), 
solid line: $\cos \phi = 1/2$, 
long-dashed line: $\cos \phi = 0$, 
long-dashed-dotted line: $\cos \phi = -1/2$. 
%


\newpage
\begin{center}
\epsfysize=10cm
\hfil\epsfbox{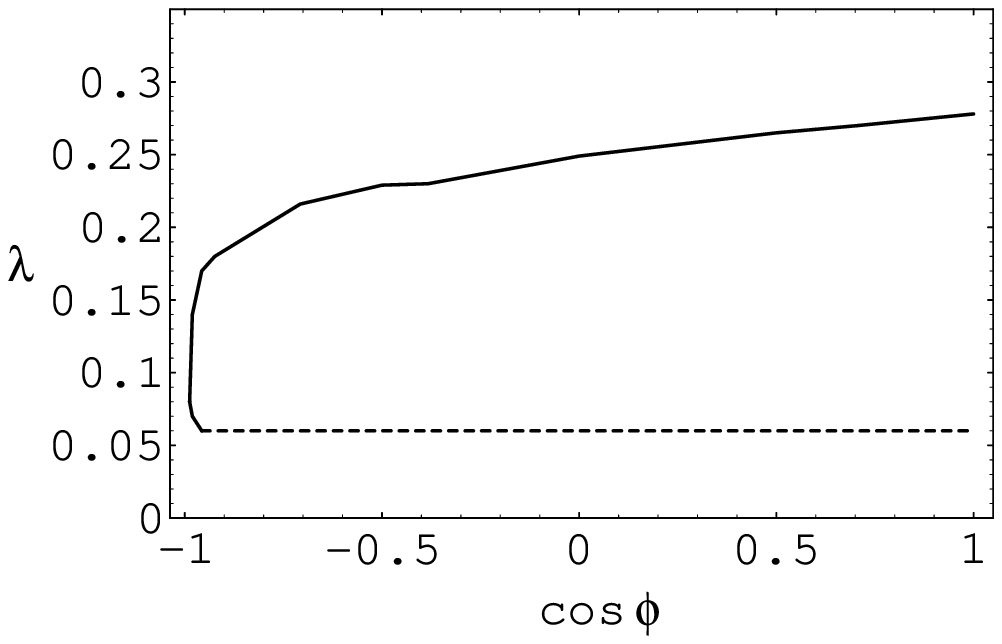}\hfill
\end{center}
\vspace{-3cm}
\begin{center}
{}Fig.1  \vspace{1cm}\\
\end{center}

\begin{center}
\epsfysize=10cm
\hfil\epsfbox{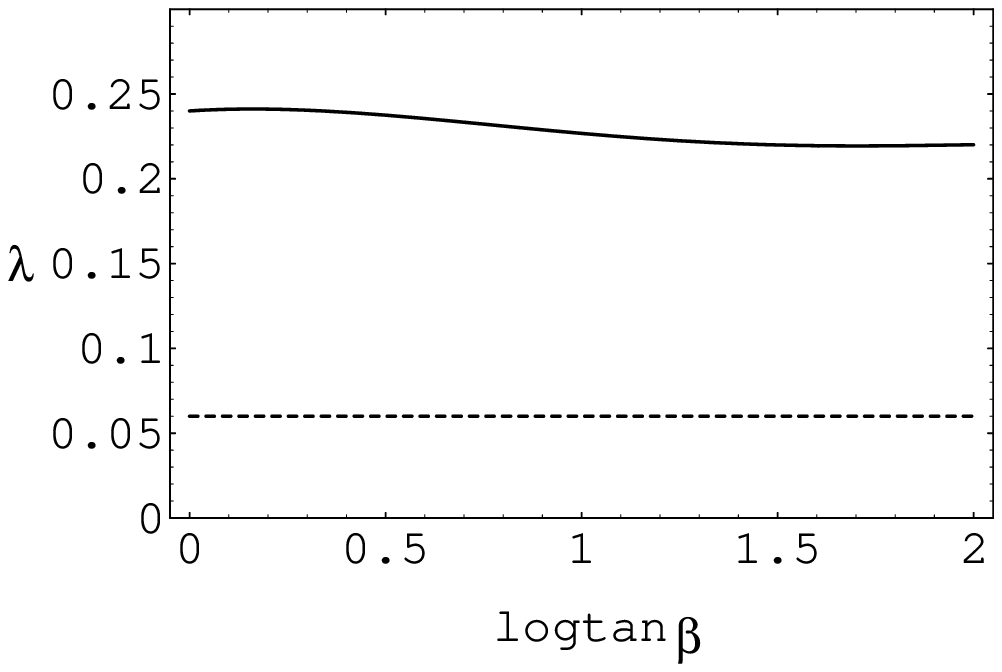}\hfill
\end{center}
\vspace{-2cm}
\begin{center}
{}Fig.2
\end{center}

\vspace*{5cm}
\begin{center}
\epsfysize=10cm
\hfil\epsfbox{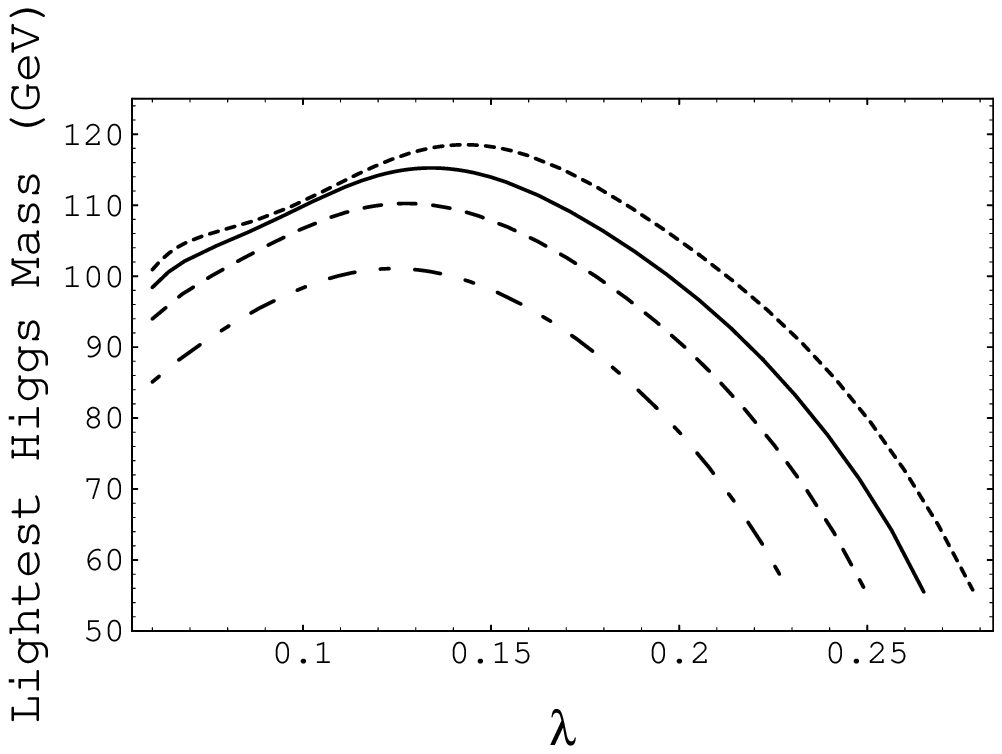}\hfill
\end{center}
\vspace{-2cm}
\begin{center}
{}Fig.3
\end{center}

\end{document}